\documentclass[12pt]{iopart}

\usepackage{graphicx}
\usepackage{bm}

\begin{document}
\title[Differences in the effects of turns and constrictions on resistive response]{
Differences in the effects of turns and constrictions on the resistive response in current-biased
superconducting wire after single photon
absorption.}
\author{A N Zotova and D Y Vodolazov}
\address{Institute for Physics of Microstructures, Russian Academy of
Sciences, 603950, Nizhny Novgorod, GSP-105, Russia}
\ead{zotova@ipm.sci-nnov.ru}

\begin{abstract}

We study how turns and constrictions affect the resistive response
of the superconducting wire after instant in time and local in
space heating, which models the absorption of the single photon by
the wire. We find that the presence of constriction favors
detection of photons of various energies but the presence of turn
increases only ability to detect relatively 'low' energy photons.
The main reason is that in case of constriction the current
density is increased over whole length and width of the
constriction while in case of the turn the current density is
enhanced only near the inner corner of the turn. It results in
inhomogeneous Joule heating near the turn and worsens the
conditions for appearance of the normal domain at relatively small
currents when the 'high' energy photons already could create
normal domain in straight part of the wire. We also find that the
amplitude of the voltage pulse depends on the place where the
photon is absorbed. It is the smallest one when photon is absorbed
near the turn and it is the largest one when photon is absorbed
near the constriction. This effect comes from the difference in
resistance of constriction and the turn in the normal state from
the resistance of the rest of the wire.

\end{abstract}

\pacs{85.25.Oj, 85.25.Pb, 07.57.Kp}
\section{Introduction}
Superconducting nanowire single-photon detectors (SNSPD) have wide
range of applications due to high sensitivity, reliability and
good time resolution \cite{review}. Basic element of such a
detector is a narrow long thin superconductive wire in a shape of
meander biased by the current close to the critical one. Recently
problem of influence of the turn in superconducting meander on the
detection efficiency attracted a lot of attention \cite{Clem,
Henrich, Hortensius, Akhlaghi, Berdiyorov}. Indeed, due to current
concentration near the turn the local current density is maximal
there. Consequently, current density achieves depairing value near
the turn at lower bias current, than in the straight wire, and
this leads to decrease of the critical current $I_c$ of the
meander as compared with a straight wire. As a result the
detecting efficiency of relatively 'low' energy photons decreases,
because it essentially depends on proximity of the critical
current to the depairing current (our definition of 'low' and
'high' energy photons see in \cite{self_ref}).

Suppression of critical current $I_c$ by turn in framework of
London model was calculated in \cite{Clem}, and experimentally
this effect was investigated in
\cite{Henrich,Hortensius,Akhlaghi}. Comparison of theory with an
experiment demonstrated qualitative agreement (the decrease of
$I_c$ with decrease of the angle of the turn and/or increase of
curvature of the turn), however quantitatively theory predicts
stronger suppression of $I_c$ than it was observed in the
experiment \cite{Henrich,Hortensius,Akhlaghi} (note that
calculations based on Ginzburg-Landau theory \cite{Henrich} gives
smaller suppression of $I_c$ than London model does). In recent
theoretical work \cite{Berdiyorov} resistive response of the wire
with the turn after photon absorption was studied and authors
found only weak effect of the turn on photon detection ability. In
our work we demonstrate, that result of Ref. \cite{Berdiyorov} is
correct only for 'high' energy photon, whereas effect is opposite
for 'low' energy photon.

In our work we also study how a narrowing (variation of the
cross-section of the wire - constriction) affects resistive
response of the superconductive nanowire after photon absorption.
Our interest to this problem was attracted by the experiments
where it was found, that voltage pulses after photon absorption
have different amplitudes \cite{Haas, Kitaygorsky}, and that with
{\it decrease} of energy of the incident photon the average
amplitude of the voltage pulse {\it increases} \cite{Haas,
Kitaygorsky}. Authors of Ref.\cite{Kitaygorsky} suggested, that
the last effect can be caused by the local inhomogeneities (for
example constrictions) of the superconducting wire. Our numerical
results confirm this suggestion and as a side effect we also find
that constriction, in contrast to turn, favors detection both
'low' and 'high' energy photons.

\section{Model}
For numerical simulations of the dynamic response after single
photon absorption we use the model which is described in detail in
our recent paper \cite{Zotova}. Shortly, we use an approach of
effective temperature of electrons \cite{Giazotto}, which is correct when
inelastic relaxation time due to electron-electron interaction
$\tau_{e-e}$ is smaller, than inelastic relaxation time due to
electron-phonon interaction $\tau_{e-ph}$. To model the effect of
the photon we assume, that at $t=0$ in the superconductor there is
an instant heating of electrons in the spot with the radius
$R_{init}$ by $\Delta T$ which is related to the energy of the
absorbed photon

\begin{equation}
   \eta2\pi \hbar c/\lambda=\Delta T \pi R_{init}^2 d C_v.
\end{equation}
Here $\eta$ is a quantum efficiency (which determines which part
of energy of the photon goes to hot electrons), $\lambda$ is an
incident electromagnetic radiation wavelength, $\hbar$ is a Plank
constant, $c$ is a speed of light, $d$ is a thickness of the wire,
$C_v$ is a specific heat capacity (for simplicity we use normal
state heat capacity at $T=T_c$). In our previous work Ref.
\cite{Zotova} we checked that the results only slightly depends on
our choice of $R_{init}$ if it is small enough and $R_{init}^2
\Delta T=const \sim 1/\lambda$.

Further evolution of the hot spot and superconducting order
parameter in the superconducting wire is based on numerical
solution of system of equations including nonstationary
Ginsburg-Landau equation, heat conductance equation for electron
temperature and Poisson equation for electric potential (note, that in the resent work \cite{Lusche} this model is compared with other models of photon detection \cite{Semenov1, Bulaevskii} and it demonstrated relatively good agreement with an experiment). This system is supplemented by the equation, which takes into account,
that the superconducting wire has finite kinetic inductance $L_k$
and a parallel connected shunting resistance $R_{shunt}$. Because
of presence of finite kinetic inductance $L_k$ in the case when
$I_s$ changes in time there is an additional voltage drop $\sim$
$L_k$(d$I_s$/dt) and voltage drop via superconductor $V_s$ is not
equal to voltage drop via shunt $V_{shunt}$ (see Eq. (5) in
\cite{Zotova} and Fig. 1).
\begin{figure}[hbtp]
\centering
\includegraphics[width=0.4\textwidth, height=45mm]{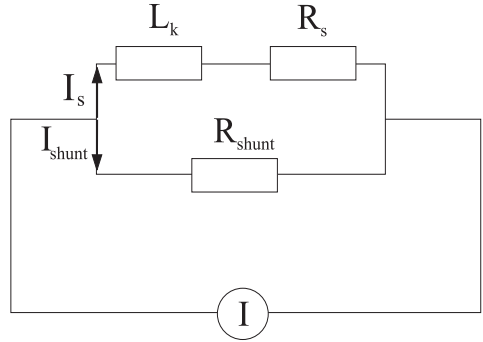}
\caption{The equivalent scheme of the superconducting detector.
The superconductor is modelled by kinetic inductance $L_k$ and
resistance $R_s$ which appeared due to absorption of the photon.
The shunt has resistance $R_{shunt}$.}
\end{figure}

\begin{figure}[hbtp]
\centering
\includegraphics[width=0.4\textwidth, height=100mm]{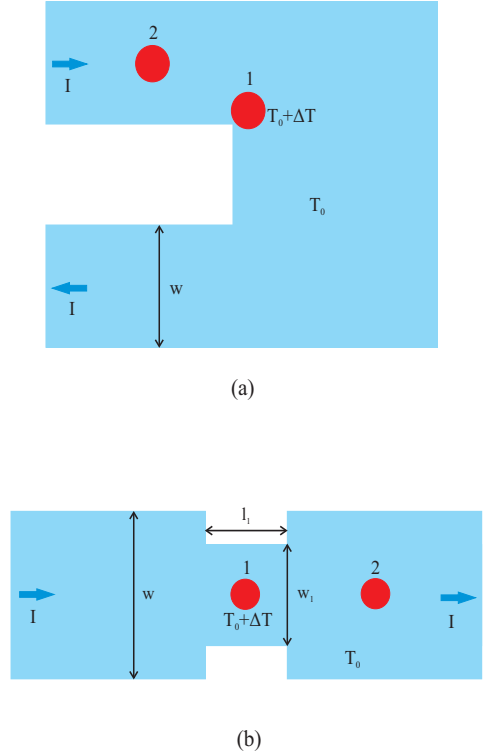}
\caption{Model geometry: (a) superconductive wire with two
$90^{\circ}$ turns (1 - photon is absorbed near the turn, 2 -
photon is absorbed far from the turn); (b) superconductive wire
with a constriction (1 - photon is absorbed in the constriction, 2
- photon is absorbed far from the constriction).}
\end{figure}

We study the wire having two $90^{\circ}$ turns (it models the
part of the meander - see figure 2(a)) and straight
superconductive wire with the constriction - see figure 2(b)
(which models variations of width of the wire - note that the
variation in the thickness of the wire produces the same effect as
variation in its width). For the parameters of the superconductor
we use the parameters roughly corresponding to NbN
\cite{Semenov,Bartolf} ($\tau_{e-e}=7$ $ps$, $\tau_{e-ph}=17$
$ps$, $L_k=0.05$ $cm$ $per$ $square$, $R_{shunt}$ = 50 $\Omega$,
critical temperature $T_c=10$ $K$, $C_v=2.4$ $mJcm^{-3}K^{-1}$,
diffusion constant $D=0.5$ $cm^2/s$, Ginzburg-Landau coherence
length at zero temperature $\xi_{GL}(0)=5$ $nm$, $d=5$ $nm$,
length of the wire $l=250$ $\mu m$, normal state resistivity
$\rho$ = 2.5 $\mu \Omega/m$) and $R_{init}$ = 7.5 $nm$. In
numerical calculations time is measured in units of $\tau_0=\pi
\hbar/8k_BT_cu\approx 0.052$ $ps$, voltage in units of
$\phi_0=\hbar/2e\tau_0\approx 6.3$ $mV$ and temperature of the
environment was fixed at $T_0 = T_c/2$.

\section{Results}

\subsection{Wire with turns}

We start with presentation of our results for the wire with two
turns (see figure 2(a)). In calculations we chose $w$ =
15$\xi_{GL}$(0) = 75 $nm$ and wire separation is equal to $w$ (the
critical current at these parameters is $I_c \simeq 0.91 I_{dep}$
at $T_0=T_c/2$), where $I_{dep}$ is a temperature dependent
depairing Ginzburg-Landau current.

\begin{figure}[hbtp]
\centering
\includegraphics[width=0.45\textwidth, height=95mm]{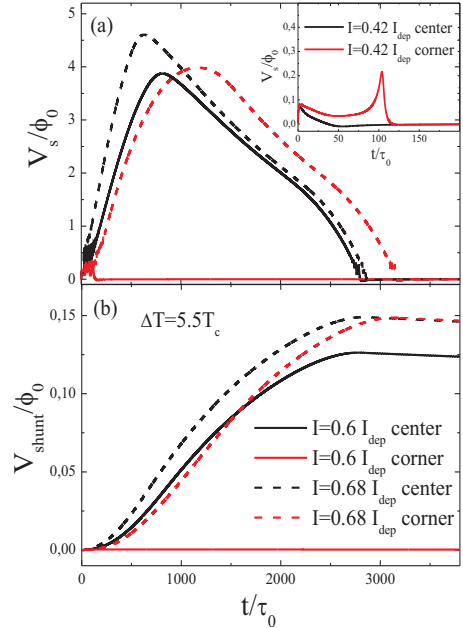}
\caption{Time dependence of voltage across (a) the superconductor
($V_s$) and across (b) the shunt ($V_{shunt}$) after absorption of
photon ($\lambda/\eta=1.7$ $\mu m$, $\Delta T$= 5.5 $T_c$) near
and far from the turn. At times larger than $3000 \tau_0$
$V_{shunt}$ decays with characteristic time
$\tau=L_k/R_{shunt}c^2\simeq 5 \cdot 10^4 \tau_0$ for our choice
of parameters. The inset shows $V_s$(t) for current $I = 0.42
I_{dep}$ for two places of photon absorption.}
\end{figure}

In figure 3 we present time dependence of voltage across the
superconductor $V_s$ and the shunt $V_{shunt}$ at different
currents with absorbed photon ($\lambda/\eta=1.7$ $\mu m$, $\Delta
T$= 5.5 $T_c$) near and far from the turn.

After photon absorption near the corner the single vortex is nucleated
at the edge of the film and passes through the film (after photon absorption
in the center of the film vortex and antivortex are nucleated simultaneously
and moves in opposite directions - in numerical calculations we could visualize
them by mapping the magnitude and phase of the calculated order parameter -
see also Figs. 3-4 in \cite{Zotova}).

From figure 3 one can see, that at relatively low current ($I = 0.6 I_{dep}$) vortices appear in series - they are seen as small peaks (noise like) in
$V_s(t)$ at $t<200 \tau_0$ after photon absorption both near the turn and far from it, but the normal domain and relatively large voltage pulse on shunt appears only for photon absorbed in straight part of the wire. Only for $I>0.66 I_{dep}$ (which was found from numerical calculations in which we varied the current and the voltage pulse appeared at $I>0.66 I_{dep}$; in Fig. 2(b) the threshold is not shown) is a pulse observed after photon absorption near the turn (see figure 3(b)).

One also can note that the amplitude of the voltage pulse is a
little smaller when the photon is absorbed near the turn (see Fig.
3(b)). Effect becomes stronger when the length of the normal
domain becomes comparable with the length of the turn (in our
model it could be reached by decreasing $R_{shunt}$ and $L_k$ or
by increasing heat removal to phonons by decreasing
$\tau_{e-ph}$). The reason for this effect is simple - in the
normal state the resistance of the wire near the turn is smaller
(because it is wider - see Fig. 2(a)) than the resistance of the
straight part of the wire and it results in difference in the
amplitudes of voltage pulses.

\begin{figure}[hbtp]
\centering
\includegraphics[width=0.45\textwidth, height=100mm]{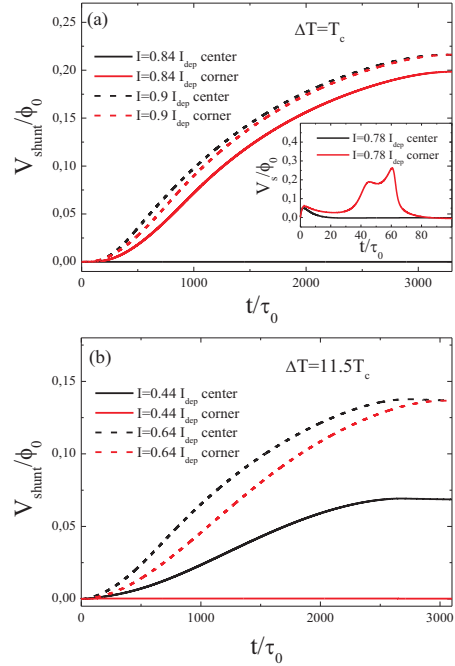}
\caption{Time dependence of voltage across the superconductor at
different currents with incident photon of different energy
($\lambda/\eta \approx$ 9.4 $\mu m$ ($\Delta T$= $T_c$ (a) and
$\lambda/\eta \approx$ 0.8 $\mu m$ ($\Delta T$= 11.5 $T_c$) (b)) near
and far from the turn of the nanowire.}
\end{figure}

In figure 4 we present results for resistive response of the
superconductor after absorption of photons of lower ($\lambda/\eta
\approx$ 9.4 $\mu m$, $\Delta T$= $T_c$) and higher energies
($\lambda/\eta \approx$ 0.8 $\mu m$, $\Delta T$= 11.5 $T_c$). The
results for 'high' energy photon (see figure 4(b)) are similar to
the results present in figure 3(a). But resistive response for
'low' energy photon is qualitatively different. The normal domain
and therefore large voltage pulse appears at smaller current when
the photon is absorbed near the turn (see figure 4(a)) and one
needs to enlarge current to observe large voltage pulse after
photon absorption far from the turn.

We explain found results as follows. The absorbed photon creates
the area with locally increased temperature of quasiparticles (or,
alternatively, increased number of quasiparticles \cite{Semenov}).
As a result the superconducting order parameter in the hot-spot
area becomes suppressed and this leads to current density
redistribution in the superconductor (it decreases in the hot-spot
area and increases around it). Nonzero $V_s$ at $t < 20 t_0$ is
associated with this process (see insets in figures 3(a) and
4(a)). The larger energy of the photon, the larger the size of the
area with partially suppressed order parameter (which could be
roughly estimated in the same way as in Refs.
\cite{Zotova,Semenov}). In Ref. \cite{Zotova} for straight
superconducting wire we show that creation of the region with
suppressed order parameter leads to nucleation of the
vortex-antivortex pair {\it inside} this region at current larger
some threshold value (which is called in \cite{Zotova} as
detecting current $I_d$). The motion of these vortices may
substantially heat the superconductor if the applied current is
large enough (larger than so called heating or retrapping current
$I_r$, which could be estimated roughly from balance of heat
dissipation and heat removal from the system - see Eq. (6) in Ref.
\cite{Zotova}) and it leads to appearance of the normal domain and
large voltage pulse via shunt. For photons of relatively large
energy (which create the large region with suppressed order
parameter) $I_d<I_r$ \cite{Zotova} and at $I \simeq I_d$ the
absorption of such a photon does not lead to large voltage pulse
despite the nucleation and motion of the vortices.

Due to intrinsically inhomogeneous current distribution (current
is concentrated near the inner corner of the turn) the order
parameter is more suppressed near the turn than far from it.
Therefore additional suppression of the order parameter due to
photon absorption favors vortex nucleation near the turn at
smaller current than photon absorption far from the turn. It is
confirmed by our numerical results for photons of both 'low' and
'high' energies (see insets in Figs. 3(a) and 4(a)) and it
coincides with the result found in Ref. \cite{Berdiyorov}.

But conditions for appearance of the normal domain is worse near
the turn than far from it. Indeed, due to inhomogeneous heating
(as a consequence of inhomogeneous current distribution even in
the normal state) and more intensive heat diffusion to the
surrounding wider superconductor the normal domain appears near
the turn at larger current than in straight part of the wire. One
also should take into account that after photon absorption near
the turn the single vortex is nucleated at the edge of the wire
and passes through the wire while after photon absorption in the
central part of straight wire vortex and antivortex are nucleated
simultaneously and moves in opposite directions. In the last case
heat dissipation is at least two times larger per unit of time
which also improves the condition for normal domain nucleation.

Retrapping or heating current $I_r$ depends also on how efficient
is the heat removal from electron subsystem to the phonons which
is governed in our model by inelastic electron-phonon relaxation
time $\tau_{e-ph}$ \cite{Zotova}. For our parameters we find that
for photons with $\lambda/\eta > 3.1$ $\mu m$ ($\Delta T < 3T_c$)
the normal domain appears at smaller current when photon is
absorbed near the turn. By changing $\tau_{e-ph}$ or using
different model for heat removal and heat dissipation one may move
this boundary. In recent work \cite{Berdiyorov} the resistive
response of the wire with the turn after absorption of the photon
with certain energy ($\Delta T$= 11.5 $T_c$ in our units) was
considered. Authors found, that the large voltage pulse appears at
smaller current with incidence of the photon in straight part of
the wire, and only at larger currents the voltage pulse appears
after photon absorption near the turn. This result coincides
qualitatively with our findings for 'high' energy photon but it is
not universal one and depends on energy of incoming photon. For
relatively 'low' energy photons the effect is opposite and the
large voltage pulse appears at smaller currents for photon acting
near the turn. There is also quantitative difference between our
findings and Ref. \cite{Berdiyorov} for the value of the voltage
pulse. Possibly, this discrepancy could be explained by the fact,
that in \cite{Berdiyorov} the value of heat removal coefficient
(which is inversely proportional to $\tau_{e-ph}$ in our model)
was by two orders of magnitude smaller than one used in our
calculations and it leaded to small heating effect in Ref.
\cite{Berdiyorov}.

\subsection{Wire with a constriction}

In this subsection we present our results for effect of the
constriction (see Fig. 2(b)) on the response of the wire after
single photon absorption. We consider the straight wire with width
$w=13\xi(0)$ = 65 $nm$, length of constriction $l_1 = w = 13
\xi(0)$ and various widths $w_1/w=0.62 \div 1$.

In the wires with width about tens of nanometers the width
variations about 10$\%$ are unlikely, but variations of thickness
could easily reach 20$\%$ (with typical thickness $d$ = 5 $nm$ it
means variation of the thickness by 1 $nm$). We consider
variations of the width because it is easier to model than the
variation of the thickness of the wire. Because both types of
constrictions lead to the same concentration of the current and
increase of the local resistance, our results could be applied to
both cases.

\begin{figure}[hbtp]
\centering
\includegraphics[width=0.5\textwidth, height=65mm]{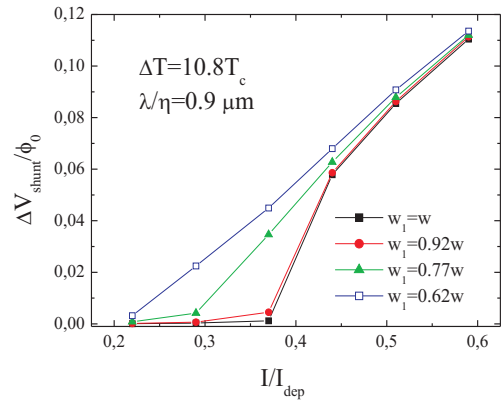}
\caption{Dependence of amplitude of the voltage pulse via shunt
resistance $V_{shunt}$ on current when a 'high' energy photon is
absorbed ($\lambda/\eta=0.9$ $\mu m$ , $\Delta T$= 10.8 $T_c$) in
the constriction with various widths.}
\end{figure}

Figure 5 demonstrates that the amplitude of the voltage pulse via
shunt increases with decreasing $w_1$ when a 'high' energy photon
is absorbed. Effect is larger at small currents when the length of
the normal domain is comparable with the length of constriction
and $\Delta V_{shunt}$ roughly scales with the width of
constriction $V_{shunt} \sim 1/w_1$. Effect becomes stronger at
large currents too when one consider longer constrictions, uses
smaller value of shunt resistance (and/or kinetic inductance) or
smaller value of $\tau_{e-ph}$ which govern the length of the
normal domain in our model.

Note that for constriction with $w_1/w \geq 0.92$ amplitude of
$V_{shunt}$ is much smaller than for narrower constrictions at $I
= 0.37 I_{dep}$. It is connected with small heating of the
superconductor and consequent nucleation of only several
vortex-antivortex pairs in the hot spot region without nucleation
of the normal domain. For constrictions with $w_1/w=0.77$ and
$w_1/w=0.62$ the normal domain appears in the wire at $I = 0.37
I_{dep}$ and it provides much larger amplitude of $V_{shunt}$.

Detection of 'low' energy photon by a constriction is similar to
case of 'high' energy photon (see figure 6(a)). At $I= 0.59
I_{dep}$ only narrowest constriction can detect 'low' energy
photon with $\lambda/\eta\approx 9.4$ $\mu m$ ($\Delta T=T_c$)
while the straight part of the wire can detect such a photons only
at $I \simeq 0.9 I_{dep}$ which is larger than the critical
current of the wire with constriction ($I_c \approx w_1/w
I_{dep}$). In this respect the detection 'ability' of constriction
is better than detection 'ability' of the turn, because it 'helps'
to detect both 'low' and 'high' energy photons. But presence of
constriction (and turn) decreases the detection ability of whole
wire because constriction/turn decreases the critical current and
hence the detection ability of the rest of the wire \cite{Kerman}.

\begin{figure}[hbtp]
\centering
\includegraphics[width=0.5\textwidth, height=95mm]{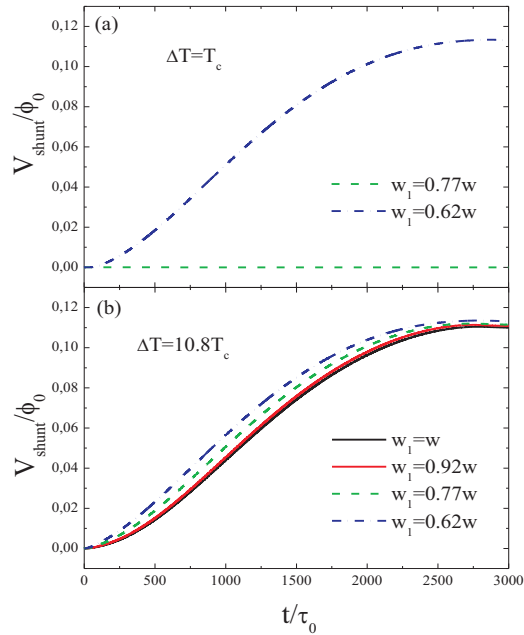}
\caption{Time dependence of $V_{shunt}$ after absorption of (a)
'low' ($\lambda/\eta=9.4$ $\mu m$ , $\Delta T$= $T_c$) and (b)
'high' ($\lambda/\eta=0.9$ $\mu m$ , $\Delta T$= 10.8 $T_c$)
energy photons in constrictions with various widths. Transport
current is fixed at $I = 0.59 I_{dep}$, just below the critical
current of the wire with the narrowest constriction $I_c \approx
0.62 I_{dep}$.}
\end{figure}

\section{Conclusion}

Effect of turn and constriction in a current biased
superconducting wire on resistive response after instant in time
and local in space heating of superconductor is studied
theoretically. In our work we assume that local heating of the
superconductor is originated from the single photon absorption by
the wire. We find that weak heating (due to absorption of 'low'
energy photon) near the turn leads to highly resistive state
(large voltage pulse) at smaller current than if the same heating
occurs in straight part of the wire and the situation is opposite
for large heating (originated from absorption of 'high' energy
photon).

We find that in contrast to the turn the presence of constriction
favors detection of both 'high' and 'low' energy photons. The main
difference between constriction and turn that in the first case
the current density is increased over whole width of the
constriction, while in case of the turn the current density is
enhanced only near inner corner of the turn and it is suppressed
in other parts of the turn which worsens the conditions for
appearance of the normal domain.

Here we have to stress that our definition of 'low' energy photon
\cite{self_ref} assumes that in absolute values it could be photon
with $\lambda=500$ $nm$ or 5 $\mu m$ depending on the width of the
meander. Apparently, in modern SNSPD with $I_c \leq 0.5
I_{dep}$ \cite{Hofherr} such a 'low' energy photons can be
detected only by parts of the meander where current density
approaches depairing current density locally (turns, local
defects) when $I \to I_c$ which leads to low intrinsic detection
efficiency (DE) of such a photons. For example in the recent paper
\cite{Marsili2} detection of 5 $\mu m$ photons with DE $\sim 1 \%
$ in SNSPD with linewidth 30 nm and 2 $\mu m$ photons with DE
$\sim 2 \% $ in SNSPD with linewidth 85 nm were experimentally
observed.

In our work we also find that the amplitude of the voltage pulse
is smaller when the photon is absorbed near the turn and it is
larger when photon is absorbed near the constriction (in
comparison with photon absorption in straight part of the wire
without constriction). We explain this effect by difference in the
resistance of the constriction, part of the wire with the turn and
the rest of the wire. Effect becomes stronger when the size of the
normal domain becomes less or comparable with the length of
constriction and length of the turn.

Due to local fluctuation (for example local increase of the temperature) in the superconducting wire the finite region with partially suppressed
order parameter will appear. In this respect effect of
fluctuation is similar to the effect of 'low' energy photon which
also creates relatively small region with suppressed
supercondutivity. Our result for 'low' energy photon shows that turns
may play decisive role in the dark counts rate if other types of
defects (for example constrictions) are absent. Indeed, the straight
part of the film in the meander is much longer than the bend region -
the reasonable estimation for their ratio is $10^3$ (it is crude estimation
from ratio of length of one line of meander $\sim 10$ $\mu m$ and one tenth
of width of single line $\sim$ 10 $nm$ - on length scale of $\sim w/10$ there
is a local enhancement of current density near the turn). But probability
for vortex entrance is proportional to $exp(-U/k_BT)$, where $U/k_BT \sim 100(1-I/I_c)$
(if one uses the London model for energy barrier for straight film -
see for example Eq. (2) in Ref. \cite{Vodolazov} and typical parameters
of NbN film with $d = 5$ $nm$ and London penetration length $\lambda_L=470$ $nm$).
But in straight part $I_c$ is larger than near the bend.
If $(1-I/I_c^{straight})=0.15$ and $(1-I/I_c^{bend})=0.05$
(which just means that $I_c^{bend} \simeq 0.89 I_c^{straight}$ and
transport current $I=0.95 I_c^{bend}$) the ratio of exponents
is $exp(10)\simeq 2\cdot 10^4$. It demonstrates the power of
the exponential factor and answers the question where the single
vortex will most probably overcome the surface barrier. Our simulations
confirm that motion of the single vortex may finally lead to appearance
of the normal domain if transport current is large enough.

Our results explain qualitatively the experimentally found finite
dispersion of amplitudes of voltage pulses observed in Refs.
\cite{Haas,Kitaygorsky}. Indeed, real superconducting meanders
have variations of width (or thickness) and turns.  Therefore the
photons with the same energy, but absorbed near or far from turn/constriction produce the voltage pulses of different
amplitude (see Fig. 3(b) and Fig. 5). Besides our results confirm
the hypothesis of work \cite{Kitaygorsky} that grow of average
amplitude of the voltage pulse with decrease of photon energy
could be explained by detection of 'low' energy photon by
inhomogeneities of the meander. It is known that with decrease of
the photon energy the detection efficiency of SNSPD drops very
fast (assuming that transport current is fixed - see for example
\cite{review}). One may suppose that voltage pulse appears only
when 'low' energy photon is incident near the relatively narrow
constriction (where current density is maximal) while parts of the
meander without constriction cannot detect such a photon. In this
case the average amplitude of the voltage pulse will be larger
(and dispersion of the amplitudes of pulses will be smaller) in
comparison with 'high' energy photon (compare Figs. 6(a) and 6(b))
because only narrow constrictions can detect 'low' energy photon.

In our simplified model we neglect heating of phonons and energy
removal to substrate which definitely affects the amplitude of the
voltage pulse via size of the normal domain \cite{Marsili,Yang}.
One may use approach of the single temperature for electrons and
phonons (like it was done in \cite{Marsili}) when the
thermoelectric processes are relatively slow and developed on time
scale much larger than both $\tau_{e-e}$ and $\tau_{e-ph}$, and
time of escape of hot phonons to substrate $\tau_{esc}$ is much
larger than max\{$\tau_{e-e}$, $\tau_{e-ph}$\}. But this condition
is definitely not valid at initial period of nucleation of the
normal domain when system decides will normal domain appear or
not. We expect that phonon heating does not influence the
condition for normal domain nucleation (in the model with
effective electron and phonon temperatures) because the
suppression of the order parameter and nucleation of first
vortices takes less than 200$\tau_0\sim$ 10 $ps$ (for parameters
of NbN in our model - see insets in Figs. 3(a) and 4(a)) which is
shorter than $\tau_{e-ph}$ in NbN and during this time one can
neglect energy transfer from electrons to phonons in two
temperature model.

At $t\gg$ 10 $ps$ one already should take into account heating of
phonons because time grow of voltage pulse is larger than
$\tau_{e-ph}$ (see Figs. 3,4) and one can use approach of Ref.
\cite{Marsili}. But it will change our results only quantitatively
- if at given material parameters and external conditions length
of the normal domain is shorter than length of constriction/turn
then one should observe noticeable variation in amplitudes of
voltage pulses and vice versus in opposite limit. Note also, that
even our simple version of heat conductance equation gives
reasonable values as for maximal value of resistance of the normal
domain ($\sim$ 1 $k \Omega$ for NbN wire with $w = 13 \xi$(0) = 65 $nm$
and $d$ = 5 $nm$) and as for rising time of $V_{shunt}$ ($t_{rise}$
$\sim$ 2000$\tau_0$ $\sim$ 100 $ps$) which are close to values
reported in the literature \cite{review,Haas} and in
\cite{Marsili} if one takes into account the difference in the
widths of the wires.

\ack This work was partially supported by the Russian Foundation
for Basic Research (project 12-02-00509) and by The Ministry of
education and science of Russian Federation (project 8686).


\begin{thebibliography}{20}
\bibitem{self_ref} In our definition the photon has 'low' energy if after absorption
it creates the hot spot (region where the order parameter is
suppressed) with radius much smaller than width of the wire. If
photon creates the hot spot with radius which is comparable with
width of the wire we call it is as 'high' energy photon. Radius of
the hot spot could be estimated, in framework of the used
theoretical model, from Eq. (16) of Ref. \cite{Zotova}. To detect
'low' energy photon one has to approach depairing (not the
critical $I_c$) current very close because only in this case the
current density near the hot spot reaches the depairing current
density and the voltage pulse will appear. For example in NbN
superconducting meanders the critical current is about $50\%$ of
depairing current \cite{Hofherr} and detection efficiency of 'low'
energy (in our sense) photons is determined by the number of
places (turns, constrictions) where the current density locally
approaches depairing value as $I\to I_c$.
\end{thebibliography}
\end{document}